\documentclass{article}
\usepackage{spconf,amsmath,graphicx}
\usepackage{subfigure}
\usepackage{multicol}
\usepackage{multirow}
\usepackage{amssymb}
\usepackage{tablefootnote}
\usepackage{multirow}
\usepackage{array}
\usepackage{marvosym}
\usepackage{xcolor} 

\title{Panoramic Image Inpainting With Gated Convolution and Contextual Reconstruction Loss}
\name{Li Yu$^{1}$, Yanjun Gao$^{1}$, Farhad Pakdaman$^{2}$, Moncef Gabbouj$^{2}$
\thanks{This work was supported by the National Natural Science Foundation of China under Grant 62002172, Business Finland AMALIA-2023, and the Marie Skłodowska-Curie grant 101022466; We acknowledge the High Performance Computing Center of Nanjing University of Information Science \& Technology for their support of this work. Corresponding author: Moncef Gabbouj (Email:moncef.gabbouj@tuni.fi).}}

\address{$^{1}$Nanjing University of Information Science and Technology, Nanjing, China 
\\$^{2}$ Tampere University, Finland}
\begin{document}
%
\maketitle
\begin{abstract}
Deep learning-based methods have demonstrated encouraging results in tackling the task of panoramic image inpainting.
However, it is challenging for existing methods to distinguish valid pixels from invalid pixels and find suitable references for corrupted areas, thus leading to artifacts in the inpainted results.
In response to these challenges, we propose a panoramic image inpainting framework that consists of a Face Generator, a Cube Generator, a side branch, and two discriminators. We use the Cubemap Projection (CMP) format as network input. The generator employs gated convolutions to distinguish valid pixels from invalid ones, while a side branch is designed utilizing contextual reconstruction (CR) loss to guide the generators to find the most suitable reference patch for inpainting the missing region.
The proposed method is compared with state-of-the-art (SOTA) methods on SUN360 Street View dataset in terms of PSNR and SSIM. Experimental results and ablation study demonstrate that the proposed method outperforms SOTA both quantitatively and qualitatively.
\end{abstract}
\begin{figure*}[t]
	\centering	
        \includegraphics[width=1\linewidth]{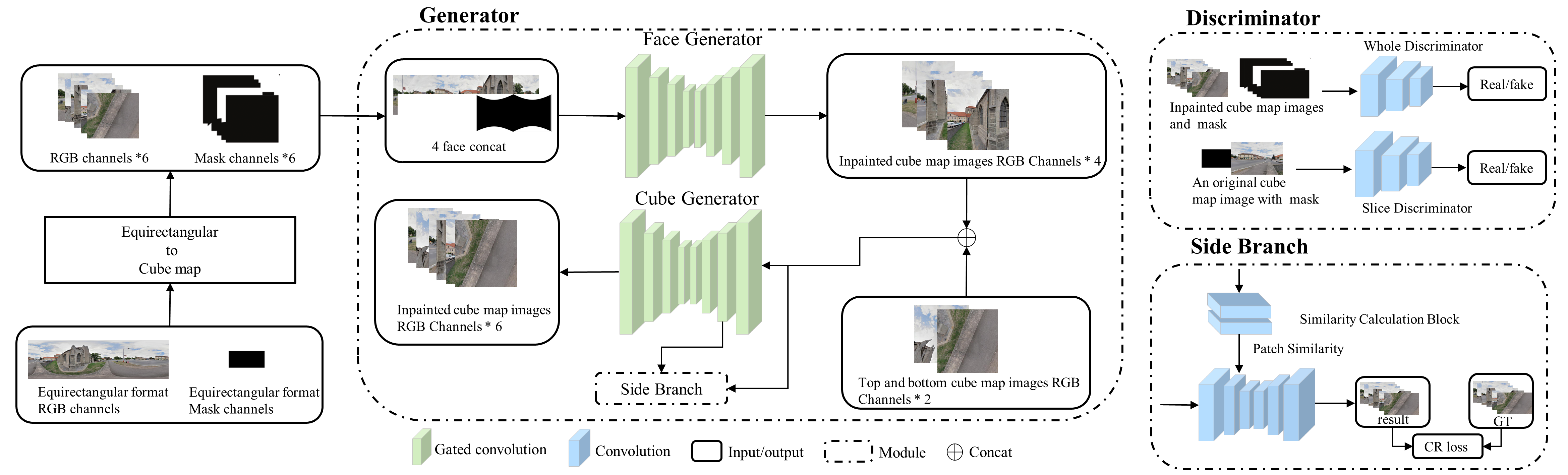}
        \caption{Our proposed panoramic image inpainting framework takes a panoramic image in CMP format as input. It uses two generators (Face Generator and Cube Generator) with gated convolution to restore the corrupted images, where the side branch guides the generator to find the most suitable reference patches. Then, each face is fed into the Slice Discriminator to judge authenticity. Finally, the six faces are simultaneously fed into the Whole Discriminator for correlation evaluation.}
	\label{FigureOne}	
 \vspace{-0.5em}
\end{figure*}
\section{Introduction}
\label{sec:intro}
Image inpainting pertains to the meticulous process of populating absent or impaired regions within an image with coherent and believable content. The goal is to seamlessly restore the image, making the inpainted regions indistinguishable from the original content. It has wide applications in areas including image restoration \cite{y-2,y-3}, distracting object removal, and image editing \cite{y-5}. With the continuous development of virtual reality (VR) technology, applications for panoramic image inpainting, such as removing unwanted objects, and filling the holes caused by stitching are expected to increase significantly.
 
Currently, the inpainting of panoramic images has yet to progress much compared to common image inpainting. Zhu et al. \cite{pii-1} presented an inpainting technique specifically designed for panoramic images, involving the projection of the panoramic image onto a sphere and subsequent completion of the missing region through matrix computations. However, this algorithm solely addresses inpainting at the bottom half of the panoramic image and is limited to images exhibiting simple patterns.
Advances in Convolutional Neural Networks (CNNs) and Generative Adversarial Networks (GANs), led to dramatic improvements in performance of image inpainting \cite{7,icassp_1,icassp_2,icassp_3}.
Akimoto et al. \cite{pii-2} designed a GAN-based inpainting method using symmetry in the panoramic image. The network uses the object's symmetry to recover the missing parts. However, this work is designed based on the ERP format, which usually leads to severe artifacts at the poles.
Han et al. \cite{pii-5} designed a panoramic image inpainting network, where the panoramic image in CMP format is used as input. This network can achieve higher-quality results by avoiding the impact of distortion in the Equirectangular Projection (ERP) format. Nevertheless, the current approach employs vanilla convolution, treating all pixels, including the ones within the corrupted image in the same way. As a consequence, it introduces visual artifacts in the result, including blurring, color inconsistencies, and evident edge responses.
To deal with this challenge, Yu et al. \cite{gated} designed gated convolution, which can be considered as a soft-gating. This learnable convolution can identify valid and invalid pixels in a corrupted image. Even at deep layers, gated convolution learns to highlight mask regions in different channels to generate better inpainting results.

An additional challenge in image inpainting arises from the existence of many potential solutions for a given missing region in images. This inherent ambiguity often gives rise to blurry or distorted regions in the final inpainted result.
To mitigate this issue, recent approaches \cite{gated,liu2019coherent,zeng2019learning,zeng2020high,xiong2019foreground} have proposed to reduce uncertainty by explicitly designating known regions as references, for inpainting missing regions. Notably, approaches such as the Contextual Attention (CA) module have emerged, which entail duplicating feature patches from a known reference area and paste them into the corresponding missing area. Nevertheless, the CA module lacks direct supervision over feature similarity and patch correspondence information, resulting in occasional selection of inappropriate patches with high weights.
Zeng et al. \cite{crfill} proposed to encourage the result of the attention-free generator to be plausible, by jointly training an auxiliary branch.

Inspired by recent works \cite{pii-5}, we propose a panoramic image inpainting network, with the panoramic image in CMP format as input. A CMP format image is composed of six independent images, often called faces. As shown in Fig. 1, we first stitch the four side faces to get a coherent and undistorted rectangular image and use the Face Generator to inpaint it. As the most important contents of the panoramic image in CMP format are on the four side faces, stitching them together can help to utilize the spatial information among them. Then, all the six faces are fed into the Cube Generator to repair the remaining up and down faces. At the same time, we introduce gated convolution in the Face Generator and Cube Generator to select useful features for reconstructing corrupted parts. Moreover, a side branch is designed as a new loss function to be jointly trained with the main network, which helps the generator to find suitable reference patches in known regions and makes the inpainting results more plausible. Next, each face of the inpainted result in CMP format is fed to a Slice discriminator to validate its authenticity, and six faces are then simultaneously fed into a Whole discriminator to judge the correlation among the six faces. In summary, our contributions can be summarized as follows:
1) We propose a GAN-based method for panoramic image inpainting, replacing the vanilla convolutions in both generators with gated convolutions to distinguish valid pixels from invalid ones.
2) We design a side branch to train jointly with the primary network, guiding the generator to find suitable reference patches for inpainting the missing area.
3) Our method achieves better performance than the SOTA method on the SUN360 dataset both qualitatively and quantitatively.

\section{Proposed Framework}
\label{sec:format}
In this paper, we propose a panoramic image inpainting network. As illustrated in Fig. 1, the inpainting architecture proposed in this paper consists of a Face Generator, a Cube Generator, a side branch (CR loss) and two discriminators. The two generators are 12-layer encoder-decoder networks with gated convolution, and the two discriminators follow the structure in paper \cite{pii-5}. Specifically, the network takes panoramic images in CMP format as input instead of ERP, because CMP format has less distortions. The network first uses the Face Generator to inpaint the stitched four side faces. Then, the output of the Face Generator is concatenated with the remaining two faces, and fed into the Cube Generator for another round of inpainting.
The side branch takes the encoder features and all faces of the inpainted image as inputs and outputs a complete image, which calculates a contextual loss. Note that the side branch is only used during training and is not needed during inference. Two discriminators are used for judgment of the output from the Cube Generator. The Slice Discriminator processes each face of the inpainted image to assess its authenticity, while the Whole Discriminator evaluates the correlation between the six faces.
\subsection{Face Generator and Cube Generator with Gated Convolution}
The four side faces of the CMP format can be stitched together to form an image with continuous content, exploiting the most important image content. 
We first restore the four side faces with the Face Generator, which consists of 6 layers of gated convolution and transposed gated convolution respectively. The gated convolution has a filter size of 4×4 and stride of 2. The transposed gated convolution has a filter size of 3×3 and stride of 1. Then, the top and bottom faces are concatenated with the four side faces output from the Face Generator, and inpainted through the Cube Generator, which has a similar structure as Face Generator. The proposed two-step generator can help the network utilize more information and produce better inpainting results.

Different from common vision tasks such as detection and classification, where all pixels of the input image are valid and can provide helpful information, in inpainting some pixels are invalid (masked). Thus, it is essential to distinguish between missing and known regions to address this issue.
To deal with this problem, we propose to replace the vanilla convolution in the two generators with gated convolutions \cite{gated}. 
The gated convolution obtains the gating value through a convolutional filter and the sigmoid function, thereby realizing a learnable soft mask update mechanism.
This mechanism ensures that pixels in the mask region of the image are not treated as valid pixels, thereby preventing invalid pixels from interfering with the encoder's capability to extract useful features from the corrupted image.
The utilization of gated convolution in the generator aligns more closely with the specific requirements of the inpainting task, thus presenting a valuable improvement over previous methods.

\subsection{Jointly Trained Side Branch}

To enhance the realism of the inpainted images, we introduce a side branch, consisting of a similarity calculation block (SCB) and a 12-layer encoder-decoder network. As shown in Fig. 1, the SCB serves as a similarity encoder, comprising two convolutional layers. The first layer employs a 5x5 filter with a stride of 1, followed by a second layer with a 3x3 filter and a stride of 1. This encoder takes the encoded feature generated by the Cube Generator as its input and yields the cosine similarity between patches.
Before the encoder feature enters the decoder, the feature patch is replaced with known region patch by using the similarity provided by SCB as weights. Then, the inpainted image is obtained through reconstruction by the decoder. Finally, the inpainted image is computed with groundtruth for CR loss. To minimize CR loss, a reference patch with highest similarity should be selected. 
During training, we optimize the network to minimize the CR loss, which involves minimizing the L1 and adversarial loss of an auxiliary result made up of image patches in known regions (more details in Loss function section). 
The network is guided to search for the optimal reference patches to approach the known image features of the minimum loss to achieve the goal of realistic content completion. 
While the CR loss indirectly impacts the generation of recovered images, it significantly enhances the network's capacity for feature learning throughout the training process.
The CR loss not only incentivizes the generator to discern the most appropriate features surrounding the missing region for reconstruction, but also enhances the visual plausibility of the inpainted part.
\subsection{Loss Function}
In order to enhance the stability of GAN training, we adopted the Wasserstein GAN (WGAN) \cite{wgangp}. The objective function of the generator in WGAN is represented by Eq. (1), while the objective function of the discriminator in WGAN is denoted by Eq. (2).
\begin{equation}
 L_{G}^{WGAN}=-E_{\hat{X} \sim P_{g}}[D(\hat{\mathrm{x}})]
\end{equation}
\begin{equation}
 L_{D}^{WGAN}=E_{\mathrm{x} \sim \mathcal{P}_{r}}[D(\mathrm{x})]
\end{equation}
where $\hat{X}$ denotes inpainted image, $X$ denotes the groundtruth image, $P_{g}$ denotes the generated data distribution, and $P_{r}$ denotes the groundtruth data distribution.

In addition, we also use WGAN gradient penalty (WGAN-gp) \cite{wgangp}, which has been modified according to the inpainting task.
The modification is defined and formulated in Eq. (3).
\begin{equation}
    L_{g p}=E_{\hat{x} \sim P_{\hat{x}}}\left(\left\|\nabla_{\hat{x}} D(\hat{\mathbf{x}}) \odot(1-\mathbf{M})\right\|_{2}-1\right)^{2} 
\end{equation}
where $\nabla$ is the gradient operation, and M is the binary mask.

The L1 losses for the mask region ($L_{1}^{mask}$) and nonmask region ($L_{1}^{non\_mask}$) are given in Eq. (4) and (5) respectively.
\begin{equation}   
L_{1}^{mask}=\left\|M \odot\left(\hat{X}-X\right)\right\| 
\end{equation}
\begin{equation}
    L_{1}^{non\_mask}=\left\|\left(1-M\right) \odot\left(\hat{X}-X\right)\right\| 
\end{equation}
The side branch uses CR loss, which is defined as:
\begin{equation}
    L_{C R}=\operatorname{ReLU}(1-D(Y \odot M+I))+\alpha\|Y-X\|_{1}
\end{equation}
where Y is the image recovered by the encoder-decoder network, and I is the incomplete image.
The overall loss for training the proposed network is given as:
\begin{equation}
   \begin{aligned}
       L_{all}=\lambda_{1}^{mask } L_{1}^{mask } +\lambda_{1}^{non\_mask } L_{1}^{non\_mask } \\ +\lambda_{D} L_{D}^{W G A N}
       +\lambda_{G} L_{G}^{W G A N}+\lambda_{C R} L_{C R}+\lambda_{gp} L_{gp}
    \end{aligned} 
\end{equation}
where, $\lambda_{1}^{mask}$, $\lambda_{1}^{non\_mask}$, $\lambda_{D}$, $\lambda_{G}$, $\lambda_{CR}$ and $\lambda_{gp}$ are the weights for mask loss, non\_mask loss, D loss, G loss, CR loss and WGAN-gp respectively. The values of weights are $\lambda_{1}^{mask}=10$, $\lambda_{1}^{non\_mask}=1$, $\lambda_{D}=1$, $\lambda_{G}=0.001$, $\lambda_{CR}=1$ and $\lambda_{gp}=10$.
\begin{figure*}
\centering
\subfigure[Input]{
\begin{minipage}[b]{0.16\linewidth}
\includegraphics[width=1\linewidth]{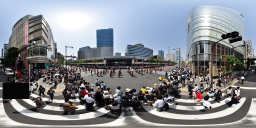}\vspace{0.5pt}
\includegraphics[width=1\linewidth]{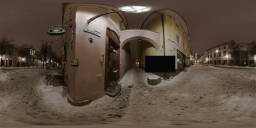}\vspace{0.5pt}
\includegraphics[width=1\linewidth]{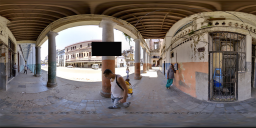}
\end{minipage}}\hspace{-0.4em}
\subfigure[ComodGAN]{
\begin{minipage}[b]{0.16\linewidth}
\includegraphics[width=1\linewidth]{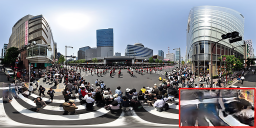}\vspace{0.5pt}
\includegraphics[width=1\linewidth]{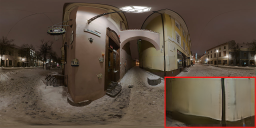}\vspace{0.5pt}
\includegraphics[width=1\linewidth]{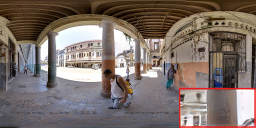}
\end{minipage}}\hspace{-0.4em}
\subfigure[PIINET]{
\begin{minipage}[b]{0.16\linewidth}
\includegraphics[width=1\linewidth]{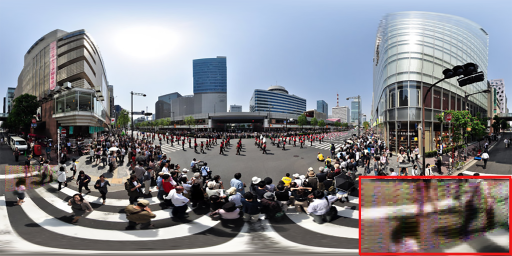}\vspace{0.5pt}
\includegraphics[width=1\linewidth]{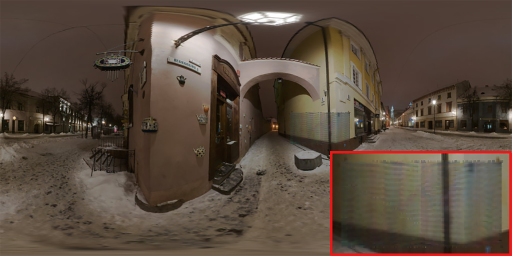}\vspace{0.5pt}
\includegraphics[width=1\linewidth]{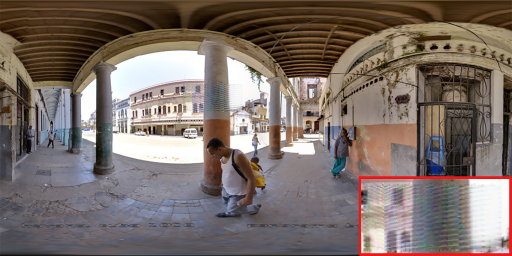}
\end{minipage}}\hspace{-0.4em}
\subfigure[PUT]{
\begin{minipage}[b]{0.16\linewidth}
\includegraphics[width=1\linewidth]{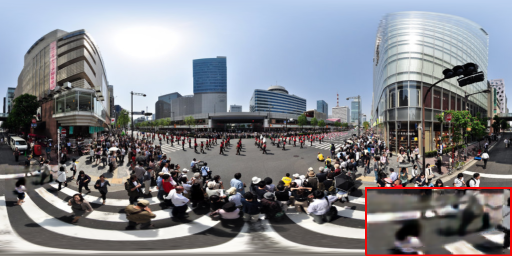}\vspace{0.5pt}
\includegraphics[width=1\linewidth]{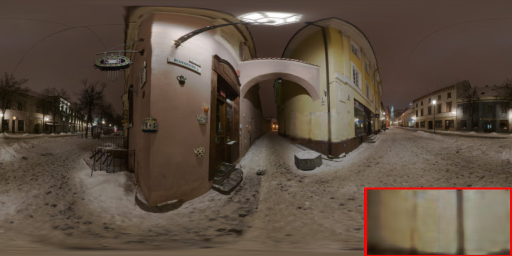}\vspace{0.5pt}
\includegraphics[width=1\linewidth]{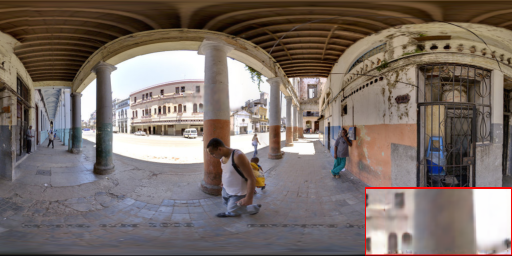}
\end{minipage}}\hspace{-0.4em}
\subfigure[Ours]{
\begin{minipage}[b]{0.16\linewidth}
\includegraphics[width=1\linewidth]{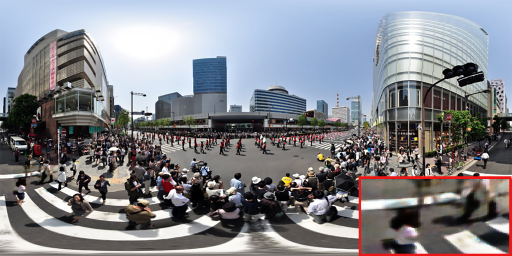}\vspace{0.5pt}
\includegraphics[width=1\linewidth]{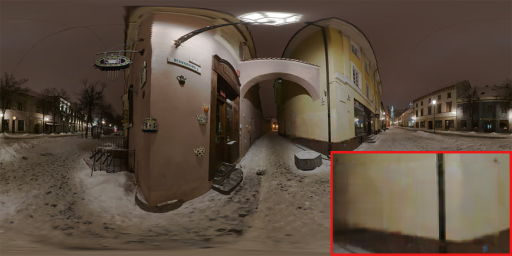}\vspace{0.5pt}
\includegraphics[width=1\linewidth]{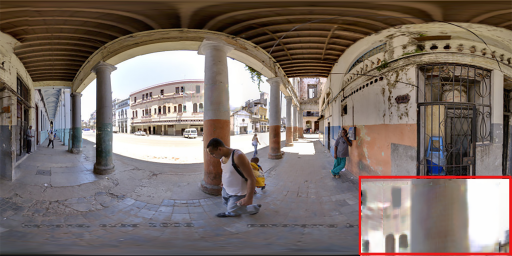}
\end{minipage}}\hspace{-0.4em}
\subfigure[GT]{
\begin{minipage}[b]{0.16\linewidth}
\includegraphics[width=1\linewidth]{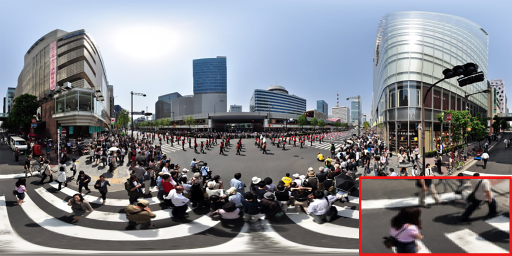}\vspace{0.5pt}
\includegraphics[width=1\linewidth]{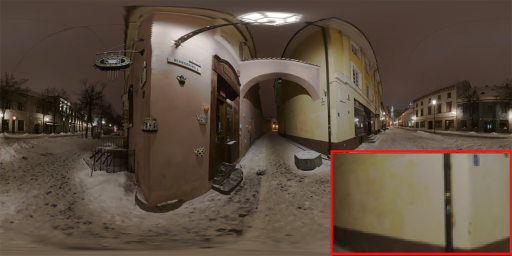}\vspace{0.5pt}
\includegraphics[width=1\linewidth]{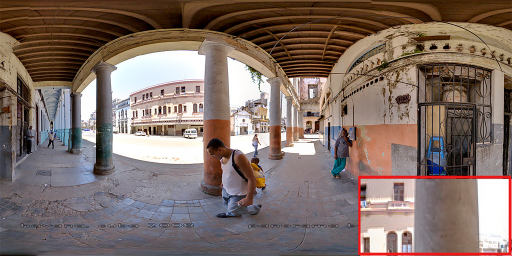}
\end{minipage}}
\caption{Qualitative comparison of the proposed method with SOTA methods. Best viewed with zoom-in.}
\vspace{-1em}
\end{figure*}
\section{Experiments And Discussion}
\label{sec: Experiments And Discussion}
In this section, we present a comprehensive evaluation of the proposed network, as well as a comparative analysis with existing approaches to assess the performance and effectiveness of our network.
We train and evaluate our model using a class of street views from the SUN360 dataset \cite{SUN360}. 
To ensure a fair comparison with SOTA methods, we conduct training and testing using the source code made available by the authors. The ERP format panoramic images have a resolution of 512×1024, while the CMP format panoramic images have a resolution of 256×256.
The proposed method was trained using PyTorch v1.7 and CUDA v10.2 on a Tesla V100 1 × 32GB. The learning rate is set to 0.0004 and batch size is 2.

\subsection{Comparative Analysis}
The quantitative comparison is done with ComodGAN \cite{7}, PIINET \cite{pii-5}, and PUT \cite{put} using PSNR and SSIM.
Table 1 illustrates the comparative performance of our method and SOTA methods at different mask ratios. Our method demonstrates superior performance at all mask ratios. For SSIM, our method achieves the highest SSIM value at low and moderate ratio masks and second best at the largest ratios.
For the 0-0.1 mask ratio, our method shows significant improvement over the baselines, with the PSNR gain of 3.50 dB on average.
Moreover, we deal with very large panoramic images with a huge field of view. Hence, even a 0.1 mask ratio represents a large region, and it is expected that improvements might saturate for larger ratios, leading to closer results among competing methods. Nevertheless, the proposed method also outperforms all competing methods at large ratios. For the 0.1-0.2 mask ratio, the average PSNR improvement over baselines is 0.89 dB, and for the 0.2-0.3 mask ratio this number is 0.34 dB.
Fig. 2 shows the qualitative comparison results.
In the first row, noticeable artifacts can be observed in the inpainted images produced by the ComodGAN and PIINET. In the PUT, the human legs in the upper right corner are blurry. Conversely, our approach effectively mitigates the occurrence of such artifacts.
In the second row, the water pipe in the inpainted image by the ComodGAN is curved, the inpainted image by the PIINET contains artifacts, and the image inpainted by the PUT has blur and artifacts. 
In contrast, the water pipe in the image inpainted by our remains straight. 
\begin{table}[t]
\centering
\caption{Quantitative comparison of the proposed method with SOTA methods on the SUN360 dataset}
\vspace{5pt}
\resizebox{0.75\columnwidth}{!}{
\begin{tabular}{c|c|c|l}
\hline
Mask Ratio & Method  & PSNR   & SSIM   \\ \hline
0-0.1      & \begin{tabular}[c]{@{}c@{}}  ComodGAN \cite{7} \\   PIINET \cite{pii-5} \\ PUT\cite{put}\\Ours\end{tabular} & \begin{tabular}[c]{@{}c@{}} 43.94\\  48.89\\49.27\\  \textbf{50.87}\end{tabular} & \begin{tabular}[c]{@{}l@{}} 0.9507\\  0.9496\\0.9510\\  \textbf{0.9512} \end{tabular} \\ \hline
0.1-0.2    & \begin{tabular}[c]{@{}c@{}} ComodGAN \cite{7} \\  PIINET \cite{pii-5} \\ PUT\cite{put}\\Ours\end{tabular} & \begin{tabular}[c]{@{}c@{}}  34.17\\  35.18\\ 35.51\\ \textbf{35.84} \end{tabular} & \begin{tabular}[c]{@{}l@{}} 0.6933\\  0.7023\\0.7055\\ \textbf{0.7061} \end{tabular} \\ \hline
0.2-0.3    & \begin{tabular}[c]{@{}c@{}}  ComodGAN \cite{7} \\  PIINET \cite{pii-5} \\PUT\cite{put}\\ Ours\end{tabular} & \begin{tabular}[c]{@{}c@{}}29.10\\  28.87\\29.05\\  \textbf{29.35} \end{tabular} & \begin{tabular}[c]{@{}l@{}}\textbf{0.6818} \\ 0.6214\\ 0.6327\\ 0.6359\end{tabular} \\ \hline
\end{tabular}}
\end{table}

\begin{table}[t]
\vspace{-4pt}
\centering
\caption{Ablation study of Gated Convolution and CR loss on the SUN360 dataset}
\vspace{5pt}
\label{table}
\resizebox{0.6\linewidth}{!}{
\begin{tabular}{c|c|c}
\hline
Gated Convolution &  CR Loss & PSNR  \\ \hline
 \textbf{×}                             &   \textbf{×}       &  48.89 \\
\textbf{×}                             &  \checkmark       &  49.56 \\
 \checkmark                             &\textbf{×}       &  50.26 \\
 \checkmark                             & \checkmark       & \textbf{50.87} \\ \hline
\end{tabular}
}
\vspace{-10pt}
\end{table}
\subsection{Ablation Study}
An ablation study is conducted to validate the effectiveness of gated convolution and CR loss within our model. 
To analyze the impact of the gated convolution, we replace it with vanilla convolution and compare the results. 
Furthermore, we assess the performance of our proposed network by excluding the Side branch and the CR loss. 
Table 2 shows the results of the ablation study.
The third and fourth rows show that the proposed framework with CR loss achieves 0.61 dB PSNR improvement.
The fourth row shows that the combination of gated convolution and CR loss achieves the best results.
The findings suggest that the gated convolution and CR loss modules exhibit effectiveness for panoramic image inpainting.
\section{Conclusion}
\label{sec:Conclusion}
In this paper, we proposed a panoramic image inpainting network with gated convolutions and a side branch. The network uses panoramic images in CMP format as input to avoid distortion at poles. Gated convolutions are introduced into the generators to distinguish between valid and invalid pixels in corrupted images. The learnable CR loss encourages the generator to search for the appropriate known region as a reference to fill the missing region. Qualitative and quantitative results demonstrated that the proposed method achieves superior performance compared to SOTA. 
\bibliographystyle{IEEEbib}
\bibliography{strings,refs}

\end{document}